\documentclass[10pt]{llncs}

\usepackage{enumerate}
\usepackage{amssymb,amsmath}
 \usepackage{xypic}
  \usepackage{comment}
   \usepackage{todonotes}

\spnewtheorem{observation}[theorem]{Observation}{\bfseries}{\itshape}

\newcommand{\alg}{{\mathbb{A}}}
\newcommand{\eps}{{\varepsilon}}
\newcommand{\Ss}{{\mathcal{S}}}
\newcommand{\clos}{{\mathbf{clos}}}
\newcommand{\td}{\mathbf{td}}
\newcommand{\dist}{{\mathbf{dist}}}
\newcommand{\height}{{\mathbf{height}}}
\newcommand{\cc}{{\mathcal{C}}}
\newcommand{\Oh}{{\mathcal{O}}}
\newcommand{\Ohstar}{{\mathcal{O}^*}}

 \title{Computing Tree-depth Faster Than $2^{n}$}

\author{Fedor V. Fomin\inst{}\thanks{Supported by European Research Council (ERC) Grant "Rigorous Theory
of Preprocessing", reference 267959.} \and Archontia C. Giannopoulou\inst{}$^{*}$ \and Micha\l{} Pilipczuk\inst{}$^{*}$}
\institute{Department of Informatics, University of Bergen,\\ P.O. Box 7803, N-5020 Bergen, Norway\\
\email{\{fomin|archontia.giannopoulou|michal.pilipczuk\}@ii.uib.no}}

\authorrunning{F. V. Fomin, A. C. Giannopoulou, M. Pilipczuk}
\titlerunning{Computing Tree-depth}
\date{}
\begin{document}
\maketitle

\begin{abstract}
A connected graph   has tree-depth at most $k$ if it is   a subgraph of  the closure of a rooted tree whose height is at most $k$. We give an algorithm which for a given  $n$-vertex graph $G$,  in  time  $\Oh(1.9602^n)$   computes the tree-depth of $G$. Our algorithm is based on combinatorial results  revealing the structure of minimal rooted trees  whose closures contain  $G$. 
\end{abstract}

\section{Introduction}

The tree-depth of a graph $G$, denoted $\td(G)$, is the minimum number   $k$ such that there is  a rooted forest $F$, not necessarily a subgraph of $G$, with the following properties. 
\begin{itemize}
\item  $V(G)=V(F)$, 
\item  Every tree in $F$ is of height at most $k$, i.e. the longest path between the root of the tree and any of its leaves contains at most $k$ vertices,
\item $G$ is a subgraph of the closure of $F$, which is the graph obtained from $F$ by adding all edges between every vertex of $F$ and the vertices contained in the  path from  this vertex to the root of the tree that it belongs to.
\end{itemize} 

 This parameter has increasingly been receiving attention  since it was defined by Ne\v{s}et\v{r}il and Ossona de Mendez in~\cite{NesetrilM06} and played a fundamental role in the theory of classes of bounded expansion~\cite{NesetrilM08,NesetrilM08a,NesetrilM08b,nesodmbook}.
Tree-depth is a very natural graph parameter, and due to different applications, was rediscovered several times under different names
 as the vertex ranking number~\cite{BodlaenderDJKKMT98}, the ordered coloring~\cite{KatchalskiMS95}, and  the minimum height of an elimination tree of a graph~\cite{NesetrilM06}. 

From the algorithmic perspective, it has been known that the problem of computing tree-depth
is {\sf NP}-hard even when restricted to bipartite graphs~\cite{BodlaenderDJKKMT98,NesetrilM06}.
However, it also admits polynomial time algorithms for specific graph classes~\cite{DeogunKKM99,KloksMW98}. For example, when the input graph is
a tree its tree-depth can be computed in linear time~\cite{Schaffer89}.
Moreover, as tree-depth is closed under minors, from the results of Robertson and Seymour~\cite{RobertsonS95b,RobertsonS04}, the problem is in {\sf FPT} when parameterized by the solution size. In~\cite{BodlaenderDJKKMT98}, Bodlaender et al. showed that the computation of tree-depth is also in {\sf XP} when parameterized by treewidth. From the point of view of approximation, tree-depth can be approximated in polynomial time within a factor of $\Oh(\log^{2}n)$~\cite{BodlaenderGHK95}, where $n$ is the number of vertices of the input graph. Moreover, there is a simple approximation algorithm that, given a graph $G$, returns a forest $F$ such that $G$ is contained in the closure of $F$ and the height of $F$ is at most $2^{\td(G)}$~\cite{nesodmbook}.
Finally, it is easy to see that there exists an exact algorithm for the computation of tree-depth running in $\Ohstar(2^{n})$ time\footnote{The $\Ohstar(\cdot)$ notation suppresses factors that are polynomial in the input size.}.

We are interested in tree-depth from the perspective of exact exponential time algorithms. Tree-depth is intimately related to another two well studied parameters, treewidth and pathwidth. 
The treewidth of a graph can be defined as   the minimum taken  over all
possible completions into a chordal graph of the maximum clique size
minus one.  Similar, path-width can be defined in terms of completion to an interval graph. 
One of the equivalent definitions of tree-depth is  as the largest clique size in a completion to a  {trivially
  perfect graph}.  These  graph classes form the following chain 

\[  \textbf{trivially perfect} \subset
\textbf{interval} \subset \textbf{chordal},\] 

\noindent corresponding to the parameters tree-depth, pathwidth, and treewidth.

However, while for the computation of   treewidth and pathwidth  there exist $\Ohstar(c^{n})$, $c<2$, time algorithms   \cite{FominKTV08,FominV12,KitsunaiKKTT12,SuchanV09}, to the best of our knowledge no such algorithm for tree-depth has been known prior to our work.
In this paper, we construct the first  exact algorithm which for any input graph $G$ computes its tree-depth in time $\Ohstar(c^{n})$, $c<2$. The running time of the algorithm is  $\Ohstar(1.9602^n)$. The approach is  
 based on the structural characteristics of the  minimal forest that
defines the tree-depth of the graph.

The rest of the paper is organized as follows. In Section~\ref{sec:prlmtd} we give some basic definitions and preliminary combinatorial results on the minimal trees for tree-depth and in Section~\ref{sec:algtd}, based on the results from Section~\ref{sec:prlmtd}, we present the $\Oh(1.9602^n)$ time algorithm for tree-depth. Finally, in Section~\ref{sec:concop} we conclude with open problems.

\section{Minimal Rooted Forests for Tree-depth}
\label{sec:prlmtd}

\subsection{Preliminaries}
For a  {graph} $G=(V,E)$, we use  $V(G)$ to denote $V$ and $E(G)$ to denote $E$.
If $S\subseteq V(G)$ we denote by $G\setminus S$ the graph obtained from $G$ after removing the vertices of $S$. In the case where $S=\{u\}$, we abuse notation and write $G\setminus u$ instead of $G\setminus \{u\}$. We denote by $G[S]$ the subgraph of $G$ induced by set $S$. 
For $S\subseteq V(G)$, the {\em open neighborhood} of $S$ in $G$, $N_{G}(S)$, is the set $\{u\in G\setminus S\mid \exists v\in S: \{u,v\}\in E(G)\}$.
Again, in the case where $S=\{v\}$ we abuse notation and write $N_{G}(v)$ instead of $N_{G}(\{v\})$.
Given two vertices $v$ and $u$ we denote by $\dist_{G}(v,u)$ their distance in $G$.
We use  $\cc(G)$ to denote the set of connected components of $G$.

\subsection{Tree-depth}

A \emph{rooted forest} is a disjoint union of rooted trees. The \emph{height} of a vertex $x$ in a rooted forest $F$ is the number of vertices of the path from the root (of the tree to which $x$ belongs) to $x$ and is denoted by $\height(x,F)$. The $\text{height}$ of $F$ is the maximum height of the vertices of $F$ and is denoted by $\height(F)$. Let $x,y$ be vertices of $F$. The vertex $x$ is an \emph{ancestor} of $y$ if $x$ belongs to the path linking $y$ and the root of the tree to which $y$ belongs. The \emph{closure} $\clos(F)$ of a rooted forest $F$ is the graph with vertex set $V(F)$ and edge set $\{\{x,y\}\mid x\mbox{~is~an~ancestor~of~} y \mbox{~in~} F, x\neq y\}$. For every vertex $y$ of $F$ we denote by $P_{y}$ the unique path linking $y$ and the root of the tree to which $y$ belongs, and denote by $p(y)$ the parent of $y$ in $F$, i.e. the neighbor of $y$ in $P_y$. Vertices whose parent is $y$ are called the {\em{children}} of $y$.
We call a vertex $x$ of $F$ a {\em branching point} if $x$ is not a root of $F$ and $\deg_{F}(x)>2$ or if $x$ is a root of $F$ and $\deg_{F}(x)\geq 2$. For a vertex $v$ of a rooted tree $T$, we denote by $T_v$ the maximal subtree of $T$ rooted in $v$. For example, if $v$ is the root of $T$, then $T_v=T$.

Let $G$ be a graph. The \emph{tree-depth} of $G$, denoted $\td(G)$, is the least $k\in\mathbb{N}$ such that there exists a rooted forest $F$ on the same vertex set as $G$ such that $G\subseteq \clos(F)$ and $\height(F)=k$.  Note that if $G$ is connected then $F$ must be a tree, and the tree-depth of a disconnected graph is the maximum of tree-depth among its connected components. Thus, when computing tree-depth we may focus on the case when $G$ is connected and $F$ is required to be a rooted tree. 

With every rooted tree $T$ of height $h$ we associate a sequence $(t_{1},t_{2},t_{3},\dots)$, where $t_{i}=|\{v\mid \height(v,T)=i\}|$, $i\in \mathbb{N}$, that is,
 $t_{i}$ is the number of vertices of the tree $T$ of height $i$, $i\in \mathbb{N}$. Note that since $T$ is finite, this sequence contains only finitely many non-zero values.

Let $T_{1}$ and $T_{2}$ be two rooted trees with heights $h_{1}$ and $h_{2}$, and corresponding sequences
$(t_{1}^{1},t_{2}^{1},t_{3}^{1},\dots)$ and $(t_{1}^{2},t_{2}^{2},t_{3}^{2},\dots)$, respectively.
 We say that
$T_{1}\prec T_{2}$ if and only if there exists an $i\in \mathbb{N}$ such that $t_{i}^{1}<t_{i}^{2}$ and $t_{j}^{1}= t_{j}^{2}$, for every $j>i$. Note in particular that if $h_{1}< h_{2}$, then taking $i=h_2$ in this definition proves that $T_{1}\prec T_{2}$.

\begin{definition}
Let $G$ be a connected graph. A rooted tree $T$ is {\em minimal} for $G$ 
if
\begin{enumerate}
\item\label{cnd1} $V(T)=V(G)$ and $G\subseteq \clos(T)$, and
\item\label{cnd2} there is no tree $T'$ such that $V(T')=V(G)$, $G\subseteq \clos(T')$, and $T'\prec T$.
\end{enumerate}
\end{definition}

The next observation follows from the definitions of $\prec$ and of  tree-depth.

\begin{observation}\label{obscragireogje}
Let $G$ be a connected graph and $T$ be a rooted tree for $G$ such that $V(T)=V(G)$, $G\subseteq \clos(T)$, and $\height(T)>\td(G)$.
Then there exists a rooted tree $T'$ such that $V(T')=V(G)$, $G\subseteq \clos(T')$, and $\height(T')<\height(T)$, and thus $T'\prec T$.
\end{observation}

The following combinatorial lemmata  reveal the structures of minimal trees which will be handy  in the algorithm.

\begin{lemma}\label{lem:mintrpod}
Let $T^{1}$ be a rooted tree with root $r$, $v\in V(T^{1})$, and $T^{*}$ be a rooted tree with root $r^{*}$ such that $T^{*}\prec T_{v}^{1}$. 
If $T^{2}$ is the rooted tree obtained from $T^{1}$ after considering the union of $T^{1}\setminus V(T_{v}^{1})$ with  $T^{*}$ and adding an edge between $r^{*}$ 
and $p(v)$ {\em (}if $p(v)$ exists{\em )}, then $T^{2}\prec T^{1}$.
\end{lemma}

\begin{proof}
Notice first that the claim trivially holds for the case where $v=r$ as then $T^{1}_{v}=T^{1}$ and $T^{2}=T^{*}$.
Thus, from now on we prove the claim assuming that $v\neq r$.
Let then $h$ be the height of the vertex $p(v)$ in $T^{1}$. 

As $T^{*}\prec T_{v}^{1}$, there exists an index $i$ such that the number of vertices of height $i$ in $T^{*}$ is strictly smaller than the number of vertices of height $i$ in $T_{v}^{1}$, and
for every $j>i$, the number of vertices of height $j$ is equal in both $T^{*}$ and $T^{1}_{v}$.  This  implies  that the number of vertices of height $h+i$ in $T^{2}$ is strictly smaller than the number of vertices of height $h+i$ in $T^{1}$, and for every $j>h+i$, the number of vertices of height $j$ is equal in both $T^{1}$ and $T^{2}$. Thus, we again conclude that $T^{2}\prec T^{1}$.
\qed\end{proof}

\begin{lemma}\label{mintrtdprop}
Let $G$ be a connected graph. If $T$ is a minimal tree for $G$ with root $r$ then for every $v\in V(T)$,
\begin{enumerate}
\item\label{cl1}  $G[V(T_{v})]$ is connected,
\item\label{cl2} $T_{v}$ is a minimal tree for $G[V(T_{v})]$, and
\item\label{cl3}  if $v'\in V(T_{v})$ is a branching point such that $\dist_{T_{v}}(v,v')$ is minimum then $N_{G}(v)\cap V(T_{u})\neq \emptyset$, for every child $u$ of $v'$. 
\end{enumerate}
\end{lemma}

\begin{proof}
We first prove~(\ref{cl1}). Assume in contrary that there exists a vertex $v\in V(T)$ such that the graph $G[V(T_{v})]$ is not connected. Notice that we may choose $v$ in such a way that $\dist_{T}(r,v)$ is maximum.
We first exclude the case where $v=r$. Indeed, notice that if $v=r$, then $G[V(T_{r})]=G$ is connected by the hypothesis. 
Thus, $v\neq r$. Notice also that if $v$ is a leaf of $T$ then $T_{v}$ is the graph consisting of one vertex, so it is again connected. Therefore, $v$ is not a leaf of $T$.
Let $v_{1},v_{2},\dots,v_{p}$ be the children of $v$. The choice of $v$ (maximality of distance from $r$) implies that $G[V(T_{v_{i}})]$ is a connected 
component of $G[V(T_{v})]\setminus v$, $i\in [p]$. Moreover, from the fact that $G[V(T_{v})]$ is not connected, it follows that there exists at least one $i_{0}\in [p]$ 
such that $N_{G}(v)\cap V(T_{v_{i_{0}}})=\emptyset$. Let $T'$ be the tree obtained from $T$ by removing the edge $\{v,v_{i_{0}}\}$ and 
adding the edge $\{p(v),v_{i_{0}}\}$.
Observe that $G\subseteq \clos(T')$. Moreover, notice that by construction of $T'$, we may consider $T'$ as the tree obtained from the union of $T\setminus V(T_{p(v)})$ with $T'_{p(v)}$ after adding the edge $\{p(v),p(p(v))\}$ (if $p(v)\neq r$).
It is easy to see that $T'_{p(v)}\prec T_{p(v)}$. Therefore, from Lemma~\ref{lem:mintrpod}, we end up with a contradiction to the minimality of $T$.

To prove~(\ref{cl2}), we assume in contrary that there exists a vertex $v\in V(T)$ such that $T_{v}$ is not a minimal tree for $G[V(T_{v})]$. By the hypothesis that $T$ is a minimal tree for $G$, it follows that $v\neq r$.
As $T_{v}$ is not a minimal tree for $G[V(T_{v})]$, there exists a rooted tree $T'$ with root $r'$ such that $V(T')=V(T_v)$, $G[V(T_{v})]\subseteq \clos(T')$, and $T'\prec T_{v}$.
Let now $T^{*}$ be the rooted tree obtained from the union of $T \setminus V(T_{v})$ with $T'$ after adding an edge between $p(v)$ and $r'$.
Notice then that $G\subseteq \clos(T^{*})$. Moreover, from Lemma~\ref{lem:mintrpod}, we get that $T^{*}\prec T$, a contradiction to the minimality of $T$.

We now prove~(\ref{cl3}). Let $v$ be a vertex of $T$ and $v'$ be a branching point of $T_{v}$  such that $\dist_{T_{v}}(v,v')$ is minimum, that is, $v'$ is the highest branching point in $T_{v}$. 
Assume in contrary that there exists a child $u$ of $v'$ such that $N_{G}(v)\cap V(T_{u})= \emptyset$. Let $T'$ be the tree obtained 
from $T$ by switching the position of the vertices $v$ and $v'$, where $T'=T$ if $v=v'$. Notice that $\clos(T)=\clos(T')$ and $T$ and $T'$ are isomorphic, hence $T'$ is also a minimal tree for $G$. Moreover, children of $v$ in $T'$ are exactly children of $v'$ in $T$. Observe also that if $w$ is a child of $v'$ in $T$, hence also a child of $v$ in $T'$, then $T_{w}=T_{w}'$ and $N_{G[V(T'_{v})]}(V(T'_{w}))\subseteq \{v\}$. 
As $N_{G}(v)\cap V(T_{u})= \emptyset$, we obtain that $G[V(T'_{v})]$ is not connected. However, $T'$ is a minimal tree for $G$ and therefore, from~(\ref{cl1}), $G[V(T'_{v})]$ is connected, a contradiction. This completes the proof of the last claim and of the lemma.
\qed\end{proof}

\section{Computing tree-depth}\label{sec:algtd}

\subsection{The naive DP, and pruning the space of states}

To construct our algorithm,  we need an equivalent recursive definition of tree-depth.

\begin{proposition}[\cite{NesetrilM06}]\label{dfn:rectd}
The \emph{tree-depth} of a connected graph $G$ is equal to 
\begin{equation}\label{eq:tdrec}
\td(G)=
\begin{cases}
1 & \text{if } |V(G)|=1\\
\displaystyle 1 + \min_{v \in V(G)} \max_{H\in \cc(G\setminus v)}\td(H) & \text{otherwise}
\end{cases}
\end{equation}
\end{proposition}

Proposition~\ref{dfn:rectd} already suggests a dynamic programming algorithm computing tree-depth of a given graph $G$ in $\Ohstar(2^n)$ time. Assume without loss of generality that $G$ is connected, as otherwise we may compute the tree-depth of each connected component of $G$ separately. For every $X\subseteq V(G)$ such that $G[X]$ is connected, we compute $\td(G[X])$ using (\ref{eq:tdrec}). Assuming that the tree-depth of all the connected graphs induced by smaller subsets of vertices has been already computed, computation of formula (\ref{eq:tdrec}) takes polynomial time. Hence, if we employ dynamic programming starting with the smallest sets $X$, we can compute $\td(G)$ in $\Ohstar(2^n)$ time. Let us denote this algorithm by $\alg_0$.

The reason why $\alg_0$ runs in pessimistic $\Ohstar(2^n)$ time is that the number of subsets of $V(G)$ inducing connected subgraphs can be as large as $\Oh(2^n)$. Therefore, if we aim at reducing the time complexity, we need to prune the space of states significantly. Let us choose some $\eps$, $0<\eps<\frac{1}{6}$, to be determined later, and let $G$ be a connected graph on $n$ vertices. We define the space of states $\Ss_\eps$ as follows:
$$\begin{array}{rccl}
\Ss_{\eps} & = & \{S\subseteq V(G)\ |\ & 1\leq |S|\leq \left(\frac{1}{2}-\varepsilon\right)n \text{ and } G[S] \text{ is connected, or }\\
& & & \exists X\subseteq V(G): |X|\leq \left(\frac{1}{2}-\varepsilon\right)n \text{ and } G[S]\in \cc(G\setminus X)\}.
\end{array}$$

Observe that thus all the sets belonging to $\Ss_\eps$ induce connected subgraphs of $G$. The subsets $S\in \Ss_\eps$ considered in the first part of the definition will be called of the {\em{first type}}, and the ones considered in the second part will be called of the {\em{second type}}. Note that $V(G)\in \Ss_\eps$ since it is a subset of second type for $X=\emptyset$.

\begin{lemma}\label{lem:Seps-enumeration}
If $G$ is a graph on $n$ vertices, then $|\Ss_{\eps}|=\Ohstar\left(\binom{n}{\left(\frac{1}{2}-\eps\right)n}\right)$.
Moreover, $\Ss_\eps$ may be enumerated in $\displaystyle \Ohstar\left(\binom{n}{\left(\frac{1}{2}-\varepsilon\right)n}\right)$ time.
\end{lemma}
\begin{proof}
For sets of the first type, there are at most $\displaystyle n\cdot \binom{n}{\left(\frac{1}{2}-\varepsilon\right)n}$ sets $S$ of size at most $\left(\frac{1}{2}-\varepsilon\right)n$. Moreover, one can enumerate them in $\displaystyle\Ohstar\left(\binom{n}{\left(\frac{1}{2}-\varepsilon\right)n}\right)$ time, and for each run a polynomial-time check whether it induces a connected subgraph. For the sets of the second type, we can in the same manner enumerate all the vertex sets $X$ of size at most $\left(\frac{1}{2}-\varepsilon\right)n$ in $\displaystyle\Ohstar\left(\binom{n}{\left(\frac{1}{2}-\varepsilon\right)n}\right)$ time, and for each of them take all of the at most $n$ connected components of $G\setminus X$.
\qed\end{proof}

In our algorithms we store the family $\Ss_\eps$ as a collection of binary vectors of length $n$ in a prefix tree (a trie). Thus when constructing $\Ss_\eps$ we can avoid enumerating duplicates, and then test belonging to $\Ss_\eps$ in $O(n)$ time.

We now define the pruned dynamic programming algorithm $\alg_\eps$ that for every $X\in \Ss_\eps$ computes value $\td_{*}(G[X])$ defined as follows:
\begin{equation}\label{eq:tdstarrec}
\td_*(G[X])=
\begin{cases}
1 & \text{if } |X|=1\\
\displaystyle 1 + \min_{v \in X} \max_{H\in \cc(G[X]\setminus v),\ V(H)\in \Ss_\eps}\td_*(H) & \text{otherwise}
\end{cases}
\end{equation}
We use convention that $\td_*(G[X])=+\infty$ if $X\notin \Ss_\eps$. The algorithm $\alg_\eps$ can be implemented in a similar manner as $\alg$ so that its running time is $\Ohstar(|\Ss_\eps|)$. We consider sets from $\Ss_\eps$ in increasing order of cardinalities (sorting $|\Ss_\eps|$ with respect to cardinalities takes $\Ohstar(|\Ss_\eps|)$ time) and simply apply formula (\ref{eq:tdstarrec}). Note that computation of formula (\ref{eq:tdstarrec}) takes polynomial time, since we need to consider at most $n$ vertices $v$, and for every connected component $H\in \cc(G\setminus v)$ we can test whether its vertex set belongs to $\Ss_\eps$ in $\Oh(n)$ time. 

For a set $S\in \Ss_\eps$ and $T$ being a minimal tree for $G[S]$, we say that $T$ is {\emph{covered by $\Ss_\eps$}} if $V(T_v)\in \Ss_\eps$ for every $v\in S$. The following lemma expresses the crucial property of $\td_*$.

\begin{lemma}\label{lem:actdvl1}
For any connected graph $G$ and any subset $S\subseteq V(G)$, it holds that $\td_{*}(G[S])\geq \td(G[S])$. Moreover, if $S\in \Ss_\eps$ and there exists a minimal tree $T$ for $G[S]$ that is covered by $\Ss_\eps$, then $\td_{*}(G[S])=\td(G[S])$.
\end{lemma}
\begin{proof}
We first prove the first claim by induction with respect to the cardinality of $S$. If $\td_*(G[S])=+\infty$ then the claim is trivial. Therefore, we assume that $S\in \Ss_\eps$, there exists some $r\in S$ such that $\td_*(G[S])=1+\max_{H\in \cc(G[S]\setminus r)}\td_*(H)$, and $V(H)\in \Ss_\eps$ for each $H\in \cc(G[S]\setminus r)$. By the induction hypothesis, since $|V(H)|\leq |S|$ for each $H\in \cc(G[S]\setminus r)$, we infer that $\td_*(H)\geq \td(H)$ for each $H\in \cc(G[S]\setminus r)$. On the other hand, by (\ref{eq:tdrec}) we have that $\td(G[S])\leq 1+\max_{H\in \cc(G[S]\setminus r)}\td(H)$. Therefore,
\begin{eqnarray*}
\td(G[S]) & \leq & 1+\max_{H\in \cc(G[S]\setminus r)}\td(H) \\
          & \leq & 1+\max_{H\in \cc(G[S]\setminus r)}\td_*(H) = \td_*(G[S]),
\end{eqnarray*}
and the induction step follows.

We now prove the second claim, again by induction with respect to the cardinality of $S$. Let $T$ be a minimal tree for $G[S]$ that is covered by $\Ss_\eps$. Let $r$ be the root of $T$ and let $v_1,v_2,\ldots,v_p$ be the children of $r$ in $T$. By (\ref{cl2}) of Lemma~\ref{mintrtdprop}, we have that $T_{v_i}$ is a minimal tree for $G[V(T_{v_i})]$, for each $i\in [p]$. Moreover, since $T$ was covered by $\Ss_\eps$, then so does each $T_{v_i}$. By the induction hypothesis we infer that $\td_*(G[V(T_{v_i})])=\td(G[V(T_{v_i})])$ for each $i\in [p]$, since $|V(T_{v_i})|<|S|$. Moreover, since $T$ and each $T_{v_i}$ are minimal, we have that
\begin{eqnarray*}
\td(G[S]) & = & \height(T)=1+\max_{i\in [p]}\height(T_{v_i})=1+\max_{i\in [p]}\td(G[V(T_{v_i})]) \\
          & = & 1+\max_{i\in [p]}\td_*(G[V(T_{v_i})]) \geq \td_*(G[S]).
\end{eqnarray*}
The last inequality follows from the fact that, by (\ref{cl1}) of Lemma~\ref{mintrtdprop}, $G[V(T_{v_i})]$ are connected components of $G[S]\setminus r$ and moreover that their vertex sets belong to $\Ss_\eps$. Hence, vertex $r$ was considered in (\ref{eq:tdstarrec}) when defining $\td_*(G[S])$. We infer that $\td(G[S])\geq \td_*(G[S])$, and $\td(G[S])\leq \td_*(G[S])$ by the first claim, so $\td_*(G[S])=\td(G[S])$.
\qed\end{proof}

Lemma~\ref{lem:actdvl1} implies that the tree-depth is already computed exactly for all connected subgraphs induced by significantly less than half of the vertices.

\begin{corollary}\label{cor:underhalf}
For any connected graph $G$ on $n$ vertices and any $S\in \Ss_\eps$, if $|S|\leq (\frac{1}{2}-\eps)n$, then $\td_{*}(G[S])=\td(G[S])$.
\end{corollary}
\begin{proof}
If $T$ is a minimal tree for $G[S]$, then for every $v\in V(T)$, $G[V(T_v)]$ is connected by (\ref{cl1}) of Lemma~\ref{mintrtdprop}, and $|V(T_{v})|\leq (\frac{1}{2}-\eps)n$. Hence, for every $v\in V(T)$ it holds that $V(T_{v})\in \Ss_\eps$ and the corollary follows from Lemma~\ref{lem:actdvl1}.
\qed\end{proof}

Finally, we observe that for any input graph $G$ the algorithm $\alg_\eps$ already computes the tree-depth of $G$ unless every minimal tree for $G$ has a very special structure. Let $T$ be a minimal tree for $G$. A vertex $v\in V(G)$ is called {\em{problematic}} if (i) $|V(T_v)|>(\frac{1}{2}-\eps)n$, and (ii) $|V(P_{p(v)})|>(\frac{1}{2}-\eps)n$. We say that a minimal tree $T$ for $G$ is {\em{problematic}} if it contains some problematic vertex.

\begin{corollary}\label{cor:problematic}
For any connected graph $G$, if $G$ admits a minimal tree that is not problematic, then $\td_*(G)=\td(G)$.
\end{corollary}
\begin{proof}
We prove that any minimal tree $T$ for $G$ that is not problematic, is in fact covered by $\Ss_\eps$. Then the corollary follows from Lemma~\ref{lem:actdvl1}.

Take any $v\in V(G)$; we need to prove that $V(T_v)\in \Ss_\eps$. First note that $G[V(T_{v})]$ is connected by~(\ref{cl1}) of Lemma~\ref{mintrtdprop}. Hence if $|V(T_{v})|\leq \left(\frac{1}{2}-\eps\right)n$, then it trivially holds that $V(T_{v})\in \Ss_\eps$ by the definition of $\Ss_\eps$. Otherwise we have that $|V(P_{p(v)})|\leq \left(\frac{1}{2}-\eps\right)n$, since $v$ is not problematic. Note then that $N_{G}(V(T_{v}))\subseteq V(P_{p(v)})$ and so $G[V(T_{v})]$ is a connected component of $V(G)\setminus V(P_{p(v)})$. Consequently, $V(T_{v})$ is a subset of second type for $X=V(P_{p(v)})$.\qed
\end{proof}

\subsection{The algorithm}

Corollary~\ref{cor:problematic} already restricts cases when the pruned dynamic program $\alg_\eps$ misses the minimal tree: this happens only when all the minimal trees for the input graph $G$ are problematic. Therefore, it remains to investigate the structure of problematic minimal trees to find out, if some problematic minimal tree could have smaller height than the one computed by $\alg_\eps$.

\begin{figure}[!h]
\begin{center}
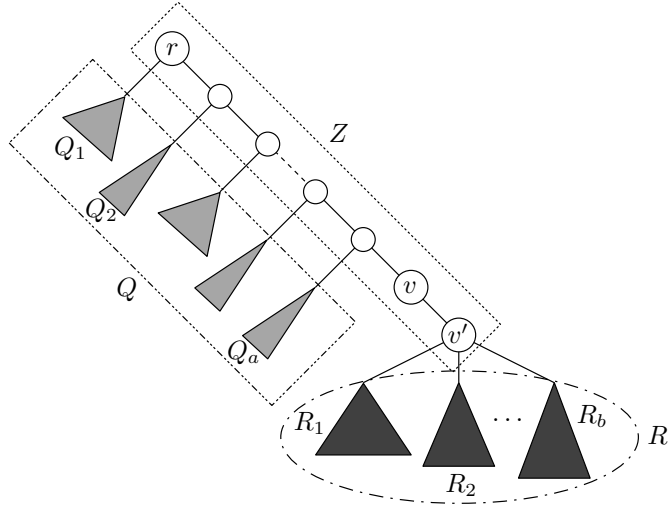
\caption{A problematic minimal tree $T$ rooted at $r$, the problematic vertex $v$, the branching vertex $v'$ in $T_{v}$, the set $Z$ (the set consisting of white vertices),  and the sets $Q_{i}$, $i\in [a]$, and $R_{j}$, $j\in [b]$. We would like to remark here that the figure only aims to facilitate identification of the above sets and that it is possible that, for some such tree $T$, the vertices in $Z\setminus v'$ have more children or there exists a vertex $q$ in $Q$ such that $\height(q,T)>\height(r,T)$ for every vertex $r\in R$.}
\label{fig:probltr}
\end{center}
\end{figure}

Let $G$ be the input graph on $n$ vertices. Throughout this section we assume that $G$ admits some problematic minimal tree $T$. 
Let $v$ be a problematic vertex in $T$. Let moreover $v'$ be the highest branching point in $T_v$ (possibly $v'=v$ if $v$ is already 
a branching point in $T$), or $v'$ be the only leaf of $T_v$ in case $T_v$ does not contain any branching points. Let $Z=V(P_{v'})$; 
observe that since $v$ is problematic, we have that $|Z|>\left(\frac{1}{2}-\eps\right)n$. Let $Q_1,Q_2,\ldots,Q_a$ be all the subtrees of
 $T$ rooted in $N_{T}(Z\setminus \{v'\})$, that is, in the children of vertices of $Z\setminus \{v'\}$ that do not belong to $Z$, and let $R_1,R_2,\ldots,R_b$ be the subtrees of $T$ 
 rooted in children of $v'$. Note that trees $Q_1,Q_2,\ldots,Q_a,R_1,R_2,\ldots,R_b$ are pairwise disjoint, and by the definition of a 
 minimal tree we have that $N_{G}(V(Q_i)),N_{G}(V(R_j))\subseteq Z$ for any $i\in [a]$, $j\in [b]$. See Figure~\ref{fig:probltr} for reference. 

Let $Q=\bigcup_{i=1}^a V(Q_i)$ and $R=\bigcup_{j=1}^b V(R_j)$. For any problematic minimal tree $T$ and a problematic vertex $v$ in it, we say that $v$ {\em{defines}} the sets $Z,Q_1,Q_2,\ldots,Q_a,R_1,R_2,\ldots,R_b,Q,R$ in $T$.

\begin{observation}\label{obs:identifyZ}
If $b>0$, then $Z=N_{G}(V(R_j)\cup Q)$ for any $j\in [b]$.
\end{observation}
\begin{proof}
As $N_{G}(V(Q_i)),N_{G}(V(R_j))\subseteq Z$ for any $i\in [a]$, $j\in [b]$, we have that $Z\supseteq N_{G}(V(R_j)\cup \bigcup_{i=1}^{a} V(Q_i))$. We proceed to proving the reverse inclusion.

Take any $z\in Z$, and let $z'$ be the highest branching point in $T_z$; note that $z'$ is always defined since $b>0$ and thus $v'$ is a branching point. If $z'=v'$, then by (\ref{cl3}) of Lemma~\ref{mintrtdprop} we infer that $z\in N_{G}(V(R_j))$, $j\in [b]$. Otherwise, we have that if $z'\in Z\setminus \{v'\}$. Since $z'$ is a branching point, there exists some subtree $Q_i$ rooted in a child of $z'$. We can again use (\ref{cl3}) of Lemma~\ref{mintrtdprop} to infer that $z\in N_{G}(V(Q_i))$, so also $z\in N_{G}(Q)$.
\qed\end{proof}

Observe that if $b=0$, then we trivially have that $Z=V(G)\setminus Q$.

\begin{observation}\label{obs:boundQ} 
$|Q|<2\eps n$.
\end{observation}
\begin{proof}
Since $v$ is problematic, we have that $|V(P_{p(v)})|>\left(\frac{1}{2}-\eps\right)n$ and $|V(T_v)|>\left(\frac{1}{2}-\eps\right)n$. Observe also that $V(P_{p(v)})\cup V(T_v)=Z\cup R$ by the definition of $Z$. Since $V(P_{p(v)})\cap V(T_v)=\emptyset$ we have that:
$$|Z\cup R| = |V(P_{p(v)})\cup V(T_v)|>(1-2\eps)n.$$
Since $Q=V(G)\setminus (Z\cup R)$, the claim follows.
\qed\end{proof}

\begin{observation}\label{obs:boundR}
$|R|<\left(\frac{1}{2}+\eps\right)n - |Q|$.
\end{observation}
\begin{proof} Since $R=V(G)\setminus (Z\cup Q)$ and $|Z|>\left(\frac{1}{2}-\eps\right)n$, we have that
$$|R|=|V(G)\setminus (Z\cup Q)|=n-|Z|-|Q|<\left(\frac{1}{2}+\eps\right)n - |Q|.$$\qed\end{proof}

\begin{observation}\label{obs:smallcomp}
If $b>0$, then $b\geq 2$ and $\displaystyle \min_{j\in [b]} |V(R_j)|< \left(\frac{1}{4}+\frac{\eps}{2}\right)n-\frac{|Q|}{2}$.
\end{observation}
\begin{proof}
If $b>0$ then $v'$ is a branching point and it has at least two children. It follows that $b\geq 2$. For the second claim, observe that since $b\geq 2$ we have that $\min_{j\in [b]} |V(R_j)|\leq |R|/2$ and the claim follows from Observation~\ref{obs:boundR}.
\qed\end{proof}

We can proceed to the description of our main algorithm, denoted further $\alg$. Similarly as before, without loss of generality let us assume that $G$ is connected. First, the algorithm constructs the family $\Ss_\eps$ using Lemma~\ref{lem:Seps-enumeration}, and runs the algorithm $\alg_\eps$ on it. Note that these steps can be performed in time $\Ohstar\left(\binom{n}{\left(\frac{1}{2}-\varepsilon\right)n}\right)$. We can therefore assume that the value $\td_*(G[S])$ is computed for every $S\in \Ss_\eps$, and in particular for $S=V(G)$.

Now the algorithm proceeds to checking whether a better problematic minimal tree $T$ with problematic vertex $v$ can be constructed. We adopt the notation introduced in the previous paragraphs for a problematic minimal tree $T$. We aim at identifying set $Z$ and sets $V(Q_1),V(Q_2),\ldots,V(Q_a),V(R_1),V(R_2),\ldots,V(R_b)$. Without loss of generality assume that if $b>0$, then $V(R_1)$ has the smallest cardinality among $V(R_1),V(R_2),\ldots,V(R_b)$, i.e., $|V(R_1)|\leq |V(R_2)|,\ldots,|V(R_b)|$. Let then $Y=Q\cup R_1$ if $b>0$, and $Y=Q$ if $b=0$.

The algorithm branches into at most $(n+1)$ subbranches, in each fixing the expected cardinality $y$ of $Y$. Note that by Observations~\ref{obs:boundQ} and~\ref{obs:smallcomp} and the fact that $\eps<\frac{1}{6}$ we may assume that 
$$y<|Q|+\left(\frac{1}{4}+\frac{\eps}{2}\right)n-\frac{|Q|}{2}=\frac{|Q|}{2}+\left(\frac{1}{4}+\frac{\eps}{2}\right)n<\left(\frac{1}{4}+\frac{3\eps}{2}\right)n.$$
Then the algorithm branches into $\binom{n}{y}$ subbranches, in each fixing a different subset of vertices of size smaller than $y$ as the set $Y$. Note that sets $V(Q_1)$, $V(Q_2),\ldots,V(Q_a),V(R_1)$ are then defined as vertex sets of connected components of $G[Y]$. The algorithm branches further into $(n+1)$ cases. In one case the algorithm assumes that $b=0$ and therefore concludes that $Q=Y$. In other cases the algorithm assumes that $b>0$ and picks one of the components of $G[Y]$ assuming that its vertex set is $V(R_1)$, thus recognizing $Q$ as $Y\setminus V(R_1)$, i.e., the union of vertex sets of remaining components of $G[Y]$.

In the case when $b=0$ the algorithm concludes that $Z=V(G)\setminus Q$. In the cases when $b>0$, the algorithm concludes that $Z=N_{G}(Y)$ using Observation~\ref{obs:identifyZ}. Having identified $Z$, the sets $V(R_1),V(R_2),\ldots,V(R_j)$ can be recognized as vertex sets of connected components of $V(G)\setminus (Z\cup Q)$. Observations~\ref{obs:identifyZ},~\ref{obs:boundQ}, and~\ref{obs:smallcomp} ensure that for every problematic minimal tree $T$ for $G$, there will be at least one subbranch where sets $Z,V(Q_1),V(Q_2),\ldots,V(Q_a),V(R_1),V(R_2),\ldots,V(R_b)$ are fixed correctly. Observe also that in each of at most $(n+1)$ branches where $y$ has been fixed, we produced at most $(n+1)\cdot \binom{n}{y}$ subbranches. We perform also sanity checks: whenever any produced branch does not satisfy any of Observations~\ref{obs:identifyZ},~\ref{obs:boundQ},~\ref{obs:boundR} or~\ref{obs:smallcomp}, or the fact that $V(R_1)$ is a smallest set among $V(R_1),V(R_2),\ldots,V(R_j)$, we terminate the branch immediately.

The algorithm now computes $\td(G[V(Q_i)])$ and $\td(G[V(R_j)])$ for all $i\in [a]$, $j\in [b]$. Recall that by Corollary~\ref{cor:underhalf}, for any set $X\subseteq V(G)$ such that $G[X]$ is connected and $|X|\leq (\frac{1}{2}-\eps)n$, we have that $\td(G[X])=\td_*(G[X])$, and hence the value $\td(G[X])$ has been already computed by algorithm $\alg_\eps$. Since $|Q|\leq 2\eps n$ and $\eps<\frac{1}{6}$, we infer that this is the case for every set $V(Q_i)$ for $i\in [a]$, and values $\td(G[V(Q_i)])$ are already computed. The same holds for every $R_j$ assuming that $|V(R_j)|\leq (\frac{1}{2}-\eps)n$. 

Assume then that there exists some $j_0$  such that $|V(R_{j_0})|>(\frac{1}{2}-\eps)n$, i.e., we have no guarantee that the algorithm $\alg_\eps$ computed $\td(G[V(R_{j_0})])$ correctly. Note that by Observation~\ref{obs:boundR} and the fact that $\eps<\frac{1}{6}$, there can be at most one such $j_0$. Furthermore, if this is the case, then by Observation~\ref{obs:smallcomp} we have that $b\geq 2$ and $V(R_{j_0})$ cannot be the smallest among sets $V(R_1),V(R_2),\ldots,V(R_b)$; hence, $j_0\neq 1$ and $V(R_{j_0})\subseteq V(G)\setminus (Z\cup Y)$. Therefore, we must necessarily have that
$$y=|Y|\leq |V(G)|-|Z|-|V(R_{j_0})|<n-\left(\frac{1}{2}-\eps\right)n-\left(\frac{1}{2}-\eps\right)n=2\eps n,$$
and moreover
$$|V(R_{j_0})|\leq |V(G)|-|Z|-|Y|<n-\left(\frac{1}{2}-\eps\right)n -y=\left(\frac{1}{2}+\eps\right)n-y.$$
Formally, if none of these assertions   holds, the branch would be terminated by the sanity check. To compute $\td(G[V(R_{j_0})])$ we employ the naive dynamic programming routine on $G[V(R_{j_0})]$, i.e., algorithm $\alg_0$. Observe, however, that in this application we do not need to recompute the values for subsets of size at most $(\frac{1}{2}-\eps)n$, since the values for them were already computed by the algorithm $\alg_\eps$. Therefore, since
$|R_{j_0}|\leq \left(\frac{1}{2}+\eps\right)n-y$ and $\eps<\frac{1}{6}$, the application of algorithm $\alg$ takes at most $\Ohstar(\binom{(\frac{1}{2}+\eps)n - y}{(\frac{1}{2}-\eps)n})$ time.

Summarizing, for every choice of $y$ (recall that $y<\left(\frac{1}{4}+\frac{3\eps}{2}\right)n$), the algorithm produced at most $(n+1)\cdot \binom{n}{y}$ branches, and in branches with $y<2\eps n$ it could have used extra $\Ohstar(\binom{(\frac{1}{2}+\eps)n-y}{(\frac{1}{2}-\eps)n})$ time for computing values $\td(G[V(R_j)])$ whenever there was no guarantee that algorithm $\alg_\eps$ computed them correctly.

We arrive at the situation where in each branch the algorithm already identified set $Z$, sets $V(Q_1),V(Q_2),\ldots,V(Q_a),V(R_1),V(R_2),\ldots,V(R_b)$, and values $\td(G[V(Q_i)])$ and $\td(G[V(R_j)])$ for $i\in [a]$, $j\in [b]$. Note, however, that the algorithm does not have yet the full knowledge of the shape of tree $T$, because we have not yet determined in which order the vertices of $Z$ appear on the path $P_{v'}$, and thus we do not know where the trees $Q_i$ and $R_j$ are attached to this path. Fortunately, it turns out that finding an optimum such ordering of vertices of $Z$ is polynomial-time solvable.

For $i\in [a+b]$ let $M_i=Q_i$ if $i\leq a$ and $M_i=R_{i-a}$ otherwise, and let $h_i=\td(G[V(M_i)])$. Note that since $T$ is minimal, by (\ref{cl2}) of Lemma~\ref{mintrtdprop} we have that $h_i=\height(M_i)$ for each $i\in [a+b]$. Let also $Z_i=N_{G}(V(M_i))$; note that since $G\subseteq \clos(T)$, we have that $Z_i\subseteq Z$. Let $\sigma$ be an ordering of $Z$, i.e., $\sigma$ is a bijective function from $Z$ to $[|Z|]$. Finally, we define the {\em{weight}} of $\sigma$ as follows:
\begin{equation}\label{eq:val}
\mu(\sigma)=\max\left(|Z|,\max_{i\in [a+b]} \left(\max(\sigma(Z_i))+h_i\right)\right).
\end{equation}

\begin{lemma}\label{lem:correctness}
Let $G$ be the input graph, and let $Z,\{V(M_i)\}_{i\in [a+b]}$ be any partitioning of vertices of $G$ such that $Z_i=N_{G}(V(M_i))$ is a subset of $Z$ for any $i\in [a+b]$. Moreover, let $h_i=\td(G[V(M_i)])$ and for $\sigma$ being an ordering of $Z$, let $\mu(\sigma)$ be defined by~\eqref{eq:val}. Then $\td(G)\leq \mu(\sigma)$ for any ordering $\sigma$ of $Z$. However, if $G$ admits a problematic minimal tree $T$ and $Z,\{V(M_i)\}_{i\in [a+b]}$ are defined by any problematic vertex in this tree, then $\td(G)=\min_{\sigma} \mu(\sigma)$.
\end{lemma}
\begin{proof}
We first prove that $\td(G)\leq \mu(\sigma)$ for any such partitioning $\{V(M_i)\}_{i\in [a+b]}$, $Z$ of $V(G)$ and ordering $\sigma$ of $Z$. 
Construct a rooted tree $T'$ as follows. First, create a path on vertex set $Z$, where the vertices are ordered as in ordering $\sigma$ and 
$\sigma^{-1}(1)$ is the root of the tree. Then, for every $i\in [a+b]$ construct a minimal tree $T_i$ for $G[V(M_i)]$, and attach its root using 
one edge to the vertex $\sigma^{-1}(m_i)$, where $m_i=\max(\sigma(Z_i))$. Observe that every neighbor of $V(M_i)$ is before $\sigma^{-1}(m_i)$ 
in the ordering $\sigma$, and hence it follows that $G\subseteq \clos(T')$. Consequently, $\td(G)\leq \height(T')$. However, by the definition of $T'$ 
and of $\mu(\sigma)$, we have   $\height(T')=\mu(\sigma)$. Thus $\td(G)\leq \mu(\sigma)$.

We proceed to the second claim. Assume that $T$ is a problematic minimal tree for $G$ and assume that $Z,\{V(M_i)\}_{i\in [a+b]}$ are defined by any problematic vertex $v$ in this tree. We adopt the notation used for $T$ in this section. Let $\sigma_0$ be the order of vertices of $Z$ on the path $P_{v'}$. For a tree $M_i$, for $i\in [a+b]$, let $z_i\in Z$ be the parent of the root of $M_i$; hence, for $i>a$ we have $z_i=v'$. Observe that, then $\td(G)=\height(T)=\max(|Z|,\max_{ i\in [a+b]} \sigma_0(z_i)+h_i)$. Since $G\subseteq \clos(T)$, we infer that $\sigma_0(z_i)\geq \sigma_0(w)$ for any $w\in Z_i$. Hence $\height(T)\geq \mu(\sigma_0)$ by the definition of $\mu$. Consequently, $\td(G)\geq \mu(\sigma_0)$, and so $\td(G)=\min_{\sigma} \mu(\sigma)$ by the first claim.
\qed\end{proof}

We are left with the following scheduling problem. Given a set $Z$ of size at most $n$, a family number of subsets $Z_i\subseteq Z$ for $i\in [a+b]$ and corresponding integers $h_i\leq n$, we would like to compute the minimum possible $\mu(\sigma)$ among orderings $\sigma$ of $Z$. Let this problem be called {\sc{Minimum Ordering with Independent Delays}} ({\sc{MOID}}, for short).

\begin{lemma}\label{lem:moid}
{\sc{Minimum Ordering with Independent Delays}} is polynomial-time solvable.
\end{lemma}
\begin{proof}
Observe that since $|Z|\leq n$ and $h_i\leq n$, for any ordering $\sigma$ we have that $\mu(\sigma)\leq 2n$. We therefore iterate through all possible values $M$ from $|Z|$ to $2n$, and for each $M$ we check whether there exists some $\sigma$ with $\mu(\sigma)\leq M$. The first $M$ for which this test returns a positive outcome is equal to $\min_{\sigma} \mu(\sigma)$.

For a given $M$, construct an auxiliary bipartite graph $H$ with $Z$ on one side and $\{1,2,\ldots,|Z|\}$ on the other side. We put an edge between an element $z$ and an index $j$ if and only if the following holds: for every $Z_i$ to which $z$ belongs, it holds that $j+h_i\leq M$. It is easy to verify that orderings $\sigma$ of $Z$ with $\mu(\sigma)\leq M$ correspond one-to-one to perfect matchings in $H$. Indeed, if we are given an ordering $\sigma$ with $\mu(\sigma)\leq M$, then we have that for every $z\in Z$ and $Z_i$ to which $z$ belongs, it holds that $\sigma(z)+h_i\leq M$ by the definition of $\mu(\sigma)$. Hence, $\{z,\sigma(z)\}$ is an edge in $H$ and $\{\{z,\sigma(z)\}\ |\ z\in Z\}$ is a perfect matching in $H$. On the other hand, if we are given a perfect matching $\{\{z,j_z\}\ |\ z\in Z\}$ in $H$, then we may define an ordering $\sigma$ of $Z$ by putting $\sigma(z)=j_z$. Then for every $z\in Z$ and $Z_i$ to which $z$ belongs, we have that $\{z,\sigma(z)\}$ is an edge in $H$ and, consequently, $\sigma(z)+h_i\leq M$. As we chose $z$ and $Z_i$ arbitrarily, it follows that $\max_{i\in [a+b]} \left(\max(\sigma(Z_i))+h_i\right)\leq M$ and so $\mu(\sigma)\leq M$.

Therefore, to solve the {\sc{MOID}} problem it suffices to construct $H$ in polynomial time and run any polynomial-time algorithm for finding a perfect matching in $H$.
\qed\end{proof}

We remark that {\sc{Minimum Ordering with Independent Delays}} can be also solved in $\Oh(n+\sum_{i=1}^{a+b}|Z_i|)$ time using greedy arguments. Since we are not interested in optimizing polynomial factors, in the proof of Lemma~\ref{lem:moid} we used the more concise matching argument to keep the description simple. We leave finding a faster algorithm for {\sc{MOID}} to the reader as an interesting exercise.

Concluding, in every subbranch algorithm $\alg$ constructs an instance of {\sc{MOID}} and solves it in polynomial time using the algorithm of Lemma~\ref{lem:moid}. Lemma~\ref{lem:correctness} ensures that none of the values found in subbranches will be larger than $\td(G)$, and that if $G$ admits a problematic minimal tree $T$ then $\td(G)$ will be found in at least one subbranch. Therefore, by Corollary~\ref{cor:problematic} we can conclude the algorithm $\alg$ by outputting the minimum of $\td_*(G)$, computed by $\alg_\eps$, and the values returned by subbranches.

Let us proceed with the analysis of the running time of algorithm $\alg$. First, we have enumerated $\Ss_\eps$ and run the algorithm $\alg_\eps$, which took 
$$T_1(n)=\Ohstar\left(\binom{n}{\left(\frac{1}{2}-\varepsilon\right)n}\right)$$
time. Then we created a number of subbranches. For every subbranch with $y\geq 2\eps n$ we have spent polynomial time, and the number of these subbranches is bounded by $(n+1)^2\cdot \binom{n}{\left(\frac{1}{4}+\frac{3\eps}{2}\right)n}$ since $y<\left(\frac{1}{4}+\frac{3\eps}{2}\right)n$ and $\eps<\frac{1}{6}$. Hence, on these subbranches we spent
$$T_2(n)=\Ohstar\left(\binom{n}{\left(\frac{1}{4}+\frac{3\eps}{2}\right)n}\right)$$
time in total. Finally, for every subbranch with $y<2\eps n$ we have spent at most $\Ohstar(\binom{(\frac{1}{2}+\eps)n-y}{(\frac{1}{2}-\eps)n})$ time. As the number of such branches is bounded by $(n+1)\cdot \binom{n}{y}$, the total time spent on these branches is
$$T_3(n)=\Ohstar\left(\max_{y<2\eps n}\left(\binom{n}{y}\cdot \binom{(\frac{1}{2}+\eps)n-y}{(\frac{1}{2}-\eps)n} \right)\right).$$
If we now let $\eps=\frac{1}{10}$, then $T_1(n),T_2(n)=\Ohstar(\binom{n}{\frac{2}{5}n})=\Ohstar(1.9602^n)$. It can be also easily shown that for any $y<\frac{1}{5}n$, it holds that $\binom{n}{y}\cdot \binom{\frac{3}{5}n-y}{\frac{2}{5}n}=\Ohstar(1.9602^n)$. To prove this, we can use the following simple combinatorial bound: $\binom{n_1}{k_1}\cdot\binom{n_2}{k_2}\leq \binom{n_1+n_2}{k_1+k_2}$. This inequality can be proved by combinatorial interpretation as follows: every choice of $k_1$ elements from a set of size $n_1$ and of $k_2$ elements from a set of size $n_2$, defines uniquely a choice of $k_1+k_2$ elements from the union of these sets, which is of size $n_1+n_2$. Therefore, we obtain:
$$\binom{n}{y}\cdot \binom{\frac{3}{5}n-y}{\frac{2}{5}n}=\binom{n}{y}\cdot \binom{\frac{3}{5}n-y}{\frac{1}{5}n-y}\leq \binom{\frac{8}{5}n-y}{\frac{1}{5}n}\leq\binom{\frac{8}{5}n}{\frac{1}{5}n}=\Ohstar(1.828^n).$$
Consequently, $T_1(n),T_2(n),T_3(n)=\Ohstar(1.9602^n)$, and the whole algorithm runs in $\Ohstar(1.9602^n)$ time.

\section{Conclusion}\label{sec:concop}
In this work we gave the first exact algorithm computing the tree-depth of a graph faster than $\Ohstar(2^n)$. As Bodlaender et al.~\cite{BodlaenderFKKT12} observe, both pathwidth and treewidth can be reformulated as vertex ordering problems and thus computed by a simple dynamic programming algorithm similar to the classical Held-Karp algorithm in time $\Ohstar(2^n)$~\cite{HeldKarp62}. For example, computing the optimum value of treewidth is equivalent to finding an elimination ordering which minimizes the sizes of cliques created during the elimination process. As far as tree-depth is concerned, Ne\v{s}et\v{r}il and Ossona de Mendez~\cite{nesodmbook} give an alternative definition of tree-depth in terms of {\em{weak-colorings}}, which in turn are defined also via vertex orderings; however, it is unclear whether this definition can be used for an algorithm working in $\Ohstar(2^n)$ time. Interestingly enough, for many of vertex ordering problems, like Hamiltonicity, treewidth, or pathwidth, an explicit algorithm working in time  $\Ohstar(c^n)$ for some $c<2$ can be designed, see \cite{Bjorklund:2010cq,FominKTV08,SuchanV09}. On the other hand, for several other vertex permutation problems no such algorithms are known. The two natural problems to attack are (i) the computation of cutwidth, and (ii) the Minimum Feedback Arc Set in Digraph problem; see \cite{BodlaenderFKKT12,CyganLPPS11a} for definitions and details. It is known that the cutwidth of a graph can be computed in time $\Ohstar(2^{t})$, where $t$ is the size of a vertex cover in the graph~\cite{CyganLPPS11a}; thus the problem is solvable in time  $\Ohstar(2^{n/2})$ on bipartite graphs.  We leave existence of faster exponential algorithms for these problems as an open question.

\bibliographystyle{abbrv}
\bibliography{tdalgo}
\end{document}